\documentstyle[12pt,floats,epsf,aps]{revtex}

\small
\begin{document}
\newcommand{\beq}{\begin{equation}}
\newcommand{\eeq}{\end{equation}}
\renewcommand{\thefootnote}{\alph{footnote}}
\title{
{\bf Multifragmentation in Xe(50A MeV)+Sn\\
Confrontation of theory and data
 } }
\author{  Regina Nebauer$^{1,2}$, J\"org Aichelin$^1$\\
\
\
\bf{and the INDRA collaboration}\\  
\
\
{\small\textrm{
 M.~Assenard$^1$,
 G.~Auger$^3$,
 Ch.O.~Bacri$^4$,
 F.~Bocage$^5$
 R.~Bougault$^5$,
 R.~Brou$^5$,
 P.~Buchet$^6$,
 J.L.~Charvet$^6$,
 A.~Chbihi$^3$,
 J.~Colin$^5$,
 D.~Cussol$^5$,
 R.~Dayras$^6$,
 A.~Demeyer$^7$,
 D.~Dor\'e$^6$,
 D.~Durand$^5$,
 P.~Eudes$^1$,
 E.~Galichet$^7$,
 E.~Genouin-Duhamel$^5$,
 E.~Gerlic$^7$,
 M.~Germain$^1$,
 D.~Gourio$^1$,
 D.~Guinet$^7$,
 P.~Lautesse$^7$,
 J.L.~Laville$^3$,
 T.~Lefort$^5$,
 R.~Legrain$^6$,
 N.~Le Neindre$^5$,
 O.~Lopez$^5$,
 M.~Louvel$^5$,
 A.M.~Maskay$^7$,
 L.~Nalpas$^6$,c
 A.D.~Nguyen$^5$,
 M.~Parlog$^8$,
 J.~P\'{e}ter$^5$,
 A.~Rahmani$^1$,
 T.~Reposeur$^1$,
 E.~Rosato$^9$,
 F.~Saint-Laurent$^3$, 
 S.~Salou$^3$,
 J.C.~Steckmeyer$^5$,
 M.~Stern$^7$,
 G.~Tabacaru$^8$,
 B.~Tamain$^5$,
 L.~Tassan-Got$^4$,
 O.~Tirel$^3$,
 E.~Vient$^5$,
 C.~Volant$^6$
 J.P.~Wieleczko$^3$ \\
{\small\textit{
$^1$ SUBATECH, IN2P3-CNRS et Universit\'e, F-44072 Nantes Cedex 03, France.~\\
$^2$ Institute for Theoretical Physics Universit\"at Rostock, Rostock, Germany.~\\
$^3$ GANIL, CEA, IN2P3-CNRS, B.P.~5027, F-14021 Caen Cedex, France.~\\
$^4$ Institut de Physique Nucl\'eaire, IN2P3-CNRS, 91406 Orsay Cedex,
France.~\\
$^5$ LPC, IN2P3-CNRS, ISMRA et Universit\'e, F-14050 Caen Cedex, France.~\\
$^6$ DAPNIA/SPhN, CEA/Saclay, 91191 Gif sur Yvette Cedex, France.~\\
$^7$ IPN Lyon, IN2P3-CNRS et Universit\'e, F-69622 Villeurbanne Cedex, France.
\\
$^8$ Nuclear Institute for Physics and Nuclear Engineering, Bucharest,
Romania.~\\
$^9$ Dipartimento di Scienze Fisiche, Univ.~di Napoli, I-180126 Napoli,
Italy.~\\}}}}}
\maketitle

\vspace{2cm}

\begin{abstract}
We compare in detail central collisions Xe(50A MeV) + Sn, recently
measured by the INDRA collaboration, with the Quantum Molecular Dynamics
(QMD) model in order to identify the 
reaction mechanism which leads to multifragmentation. We find that QMD 
describes the data
quite well, in the projectile/target region as well as in the midrapidity zone 
where also statistical models can be and have been employed.
The agreement between QMD and data allows to use this dynamical model
to investigate the reaction in detail. We arrive at the following observations:
a) the in medium nucleon nucleon
cross section is not significantly different from the free cross section, b)
even the most central collisions have a binary character, c) most of the
fragments are produced in the central collisions and d) the simulations as well as
the data show a strong attractive in-plane flow resembling deep inelastic
collisions e) at midrapidity the results from QMD and those from statistical
model calculations agree for almost all observables with the exception of
${d^2 \sigma \over dZdE}$. This renders it difficult to extract the reaction
mechanism from midrapidity fragments only. 
According to the simulations the reaction shows a very early 
formation of fragments, even in central collisions, which pass through
the reaction zone without being destroyed. The final transverse momentum of the
fragments is very close to
the initial one and due to the Fermi motion. A heating up of the systems is not observed and hence a thermal
origin of the spectra cannot be confirmed.
\end{abstract}

\vspace{2cm}

\section{Introduction}
Why does a nucleus shatter into several (up to a dozen)
intermediate mass fragments (IMF's, $Z \ge 3$)) if hit by a projectile
nucleus? Is this only a statistical or even a thermal process and hence 
(micro)canonical
phase space models are the proper tool for its description
\cite{bon95} - \cite{QSM} or is this a dynamical process,
for example similar to the shattering of glass, 
as also conjectured \cite{AH}? 

Despite of extensive  efforts of several experimental groups 
\cite{fopi95} - \cite{rei} 
this question is not finally decided yet. 
The main reason is that, surprisingly
enough, both approaches give very similar results for at least two
key observables. If multifragmentation is a thermal process and due to the 
liquid gas phase transition, predicted by the nuclear matter Hamiltonian for a
density around a third of normal nuclear matter density,
one expects a mass yield curve of the form of a
power law $\sigma(A) \propto A^{-\tau}$. The same is true if 
multifragmentation is a process 
similar to the shattering of glass \cite{AH}. If multifragmentation is a slow process 
and the
system reaches and maintains a global equilibrium before it fragments,
the average transverse kinetic energy of the fragments is equals 3/2 kT and
independent of the fragments size.
If the opposite is true and multifragmentation is a very fast
process in which the fragments retain their initial Fermi momentum the
fragments have as well a transverse kinetic energy independent of their mass of 2/5$E_F$  
\cite{gol}, where $E_F$ is the Fermi energy of the nucleus. Hence neither the mass 
yield curve nor the average kinetic energies
of fragments allow for a distinction between the two quite different mechanisms.
Rather
one has to study more exclusive observables or many body correlations.
This requires usually high statistics $4 \pi$ experiments.

There are further complications. One expects that the reaction mechanism depends
on the impact parameter. To understand details of the
reaction or even the reaction mechanism from inclusive data has turned out
to be hopeless. Thus an effective event selection is necessary. At low beam
energies ($30A\ MeV < E_{kin} < 150A\ MeV$) such a selection is difficult because
the available phase space is very small and hence it is not easy 
to
find effective selection criteria.
The most useful event selection criteria like charged particle
multiplicity or total transverse energy require high granularity $4 \pi$
detectors specially devoted to study multifragmentation.   

These high granularity low threshold $4\pi$ detectors became available only 
recently at the GSI in Darmstadt
(Germany), at GANIL in Caen (France) and at the Michigan State University
(USA). The results from these detectors allow now for a new effort to understand
multifragmentation.

It is the purpose of this paper to compare in detail the recently obtained 
experimental results of the INDRA collaboration for the reaction 
Xe(50A MeV)  + Sn with the predictions of the Quantum Molecular Dynamics
(QMD) approach, a dynamical model suited to describe the formation of fragments.
This reaction has been chosen for two reasons. First of all, the INDRA detector
has the highest efficiency of all presently operating $4 \pi$ detectors and
therefore the results present a challenge for every theory. Secondly, at 
50A MeV the number of produced fragments has a maximum \cite{pei} and therefore it is the
proper energy to study multifragmentation. It is also an energy where the
Fermi spheres of projectile and target become separated and hence
one may expect that it is the beginning of the transitional regime between 
low energy heavy ion reactions, characterized by compound nucleus formation 
and deep inelastic collisions, and high energy heavy ion reactions,
characterized by two types of nucleons, the participants and the spectators.

We start out with a short description of the QMD model in section \ref{qmo}. 
A comparison of simulations with experiment sounds easier than it is.
In order to compare them it is necessary to employ a filter
which tells us how the detector would register the theoretical data. It has
to determine how the detector reacts if hit by 2 particles in the same reaction,
whether a particle has hit 
the active zone of one of the detectors and whether the particle has an energy
below the (detector dependent) threshold and hence is not registered. It
has as well to take into account the deceleration of the particles while 
passing through the target.
The filter is not only of importance for the prediction of single particle 
observables but as well for the proper selection of the subset of data
which are the basis of the analysis. Thus an event selection is necessary
if one wants to select a small impact parameter range.
Therefore, in section \ref{filter}, we discuss the influence of the filter on
the theoretical QMD data 
and  motivate our criterion for selecting 
the most central events. Section \ref{sipa} presents the single particle spectra for
light charged particles followed by an analysis of the global fragment observables
in section VI.  As we will see in section \ref{angd}, we can separate two angular ranges. In the
nucleus nucleus center of mass frame 
the angular distribution is flat between $\theta = 60^o$ and $\theta =  120^0$.
For smaller or larger
angles the angular distribution increases towards the beam axis. We analyze both
regions separately in sections \ref{mid} and \ref{en1}. Section \ref{bla} is
 devoted to a
discussion of other proposed criteria to select central events and we compare
the results obtained under these selection criteria with those obtained with our
choice. Finally, in section \ref{reac}, we discuss the reaction scenario which
emerges from the simulations and draw our conclusions.

\pagebreak

\section{QMD model}\label{qmo}
The QMD model is a time dependent A-body theory to simulate the time evolution 
of heavy ion reactions on an event by event basis.
It is based on a generalized variational principle. As every variational
approach it requires the choice of a test wave function $\phi$. In the QMD
approach this is an A-body wave function with 6 A time dependent parameters
if the nuclear system contains A nucleons. 

To calculate the
time evolution of the system we start out from the action

\beq
S=\int_{t_1}^{t_2} {\mathcal L}\left [ \phi,\phi^*\right ] dt
\eeq

with the Lagrange functional 
\beq
{\mathcal L}=\left\langle\Phi\left\vert i\hbar\frac{d}{dt} -
H\right\vert\Phi\right\rangle.
\eeq
The total time derivative includes the derivation with respect to the parameters. The
time evolution of the parameters is obtained by the requirement that the action is
stationary under the allowed variation of the wave function. This leads to an
Euler-Lagrange equation for each time dependent parameter.

The basic assumption of the QMD model is that a test wave function of the form 

\beq
\Phi = \prod_{i=1}^{A_T+A_P}\phi_i
\eeq
with
\beq
\phi_i (\vec{r},t) = 
\left({\frac{2 }{L\pi}}\right)^{3/4}\, e^{-(\vec{r} -
\vec{r_i}(t))^2/4L} \,e^{i(\vec{r}-
\vec{r_i}(t)) \vec{p_i}(t)}\,
e^{i p_i^2(t)t/2m}. 
\eeq

is a good approximation to the nuclear wave function. The time dependent parameters are
$\vec{r_i}(t), \vec{p_i}(t)$, L is fixed and equals 1.08 $fm^2$. Thus the
rms radius of a nucleon is about 1.8 fm and hence almost twice as large
as that obtained from electron scattering. A smaller value of L is excluded
because the nuclei would become unstable after initialization. Thus this value
of L presents the limit for a semiclassical theory. The consequences of this, longer
interaction range will be discussed along this article.

Variation yields:
\begin{equation}
\dot{\vec{r}_i} = {\frac{\vec{p}_i }{m}} + \nabla_{\vec{p}_i} \sum_j
\langle V_{ij}\rangle  = \nabla_{\vec{p}_i} \langle H \rangle
\end{equation}
\begin{equation}
\dot{\vec{p}_i} = - \nabla_{\vec{r}_i} \sum _{j\neq i} \langle
V_{ij}\rangle = -\nabla_{\vec{r}_i} \langle H \rangle
\end{equation}
with $\langle V_{ij}\rangle = \int d^3x_1\,d^3x_2\, 
\phi_i^* \phi_j^* V(x_1,x_2) \phi_i \phi_j$.
These are the time evolution equations which are solved numerically. 
Thus the variational principle reduces the time evolution of the 
$n$-body Schr\"odinger equation to the time evolution equations of 
$6 \cdot (A_P+A_T)$ parameters to which a physical meaning can be
attributed.

The nuclear dynamics of the QMD can also be translated into a
semiclassical scheme. The Wigner distribution function $f_i$ of 
the nucleon $i$ can be easily derived 
from the test wave functions (note that antisymmetrization is neglected) 
\begin{equation} \label{fdefinition}
 f_i (\vec{r},\vec{p},t) = \frac{1}{\pi^3 \hbar^3 }
 {\rm e}^{-(\vec{r} - \vec{r}_{i} (t) )^2  \frac{1}{2L} }
 {\rm e}^{-(\vec{p} - \vec{p}_{i} (t) )^2  \frac{2L}{\hbar^2}  }
\end{equation} 
and the total one body Wigner density is the sum of those of all nucleons.
The potential can be calculated with help of the wave function or of the Wigner density.
Hence the expectation value of the total Hamiltonian reads 
\begin{eqnarray} 
\langle H \rangle &=& \langle T \rangle + \langle V \rangle 
\nonumber \\ \label{hamiltdef}
&=& \sum_i \frac{p_i^2}{2m_i} +
\sum_{i} \sum_{j>i}
 \int f_i(\vec{r},\vec{p},t) \,
V^{ij}  f_j(\vec{r}\,',\vec{p}\,',t)\,
d\vec{r}\, d\vec{r}\,'
d\vec{p}\, d\vec{p}\,' \quad.
\end{eqnarray}
The baryon-baryon potential $V_{ij}$ consists of the real part of the Br\"uckner 
$G$-Matrix which is supplemented by an effective Coulomb interaction
between the charged particles. The former can be further subdivided into 
a part containing the contact Skyrme-type interaction only and a contribution
due to a finite range Yukawa-potential.
$V^{ij} $
consists of 
\begin{eqnarray}
V^{ij} &=& G^{ij} + V^{ij}_{\rm Coul} \nonumber \\
       &=& V^{ij}_{\rm Skyrme} + V^{ij}_{\rm Yuk} + 
           V^{ij}_{\rm Coul} \nonumber \\
       &=& t_1 \delta (\vec{x}_i - \vec{x}_j) + 
           t_2 \delta (\vec{x}_i - \vec{x}_j) \rho^{\gamma-1}(\vec{x}_{i}) +
           t_3 \frac{\hbox{exp}\{-|\vec{x}_i-\vec{x}_j|/\mu\}}
               {|\vec{x}_i-\vec{x}_j|/\mu}  \\
       & & +  \frac{Z_i Z_j e^2}{|\vec{x}_i-\vec{x}_j|} \nonumber
\end{eqnarray}
The range of the Yukawa-potential is chosen as 1.5 fm.
$Z_i,Z_j$ are the effective charges ${{Z_p}\over{N_p}},{{Z_t}\over{N_t}}$
of the baryons $i$ and $j$. 
The real part of the Br\"uckner G-matrix is density dependent, which is
reflected in the expression for $G^{ij}$.
The expectation value of $G$ for the nucleon $i$ 
is a function of the interaction density 
$\rho_{\rm int}^i$. 
\begin{equation} \label{rhoint}
\rho_{\rm int}^i(\vec{r_i}) = \frac{1}{(\pi L)^{3/2}} 
\sum_{j \neq i} {\rm e}^{\displaystyle 
-(\vec{r_{i}}-\vec{r_{j}})^2/L }.
\end{equation}
Note that the interaction density 
has twice the width of the single particle density.

The imaginary part of the G-matrix acts like a collision 
term. In the QMD simulations we restrict ourselves to binary collisions 
(two-body level). The collisions are performed in a point-particle sense in a 
similar way as in VUU or in cascade calculations: Two particles may collide if
they come closer than $r=\sqrt{\sigma/\pi}$ where $\sigma$ is a parametrization of the
free NN - cross section. 
A collision does not take place if the
final state phase space of the scattered particles is already occupied 
by particles of the same kind (Pauli blocking).

Neglecting antisymmetrization is the most drastic approximation of the model. 
Thus, all
properties related to shell structures cannot be accounted for.
The binding energy per nucleon follows the Weizs\"acker mass formula. Hence,
 small
fragments which show a large deviation from that formula cannot be reproduced
quantitatively. The initial values of the
parameters are chosen in that way that the nucleons give proper densities and
 momentum
distributions of the projectile and target nuclei.

Fragments are determined in this model by a minimum spanning tree procedure.
At the end of the reaction all those nucleons are part of a fragment which
have a neighbor in a distance less the $r_{frag}= 3fm$. After 200 fm/c the
fragment multiplicity remains stable because only bound nucleons remain 
together whereas the others separate in the expanding system. The result 
is also
not sensitive to $r_{frag}$ in a reasonable interval for $r_{frag}$.

For further details of the QMD- model we refer to ref.\cite{aic,hartn}

To compare the QMD simulations with
experimental data as realistic as possible 
we built up a data base of about 60 000 QMD events over a large impact parameter
range. We have chosen a soft equation of state.


\section{The experimental filter} \label{filter}

We start our comparison of the INDRA results with QMD calculations with
a discussion of the filter. To filter simulations of heavy ion collisions
is an absolute must if one would like to compare experiment
and calculation in a quantitative way. A filter routine is a very complicated
program. It has not only to take into account which particles are not
registered because they hit detector frames or disappear in the beam pipe.
It also has to answer the question how the detector would react if by
accident two particles enter the same detector and it has to take into account
how target like fragments with a very low energy pass through the solid 
target material, i.e. whether they arrive at all at the detectors.
In addition, it has to
reproduce exactly the energy thresholds of the different detectors, which is
of special importance for these low energy reactions where the thresholds
are not far away from the maxima of the fragment energy distributions.
A minimum of 4 charged products is required in both the filtered QMD events as well as 
for the INDRA data.

\begin{figure}[h]
\vspace{-3cm}
\epsfxsize=16.cm
$$
\epsfbox{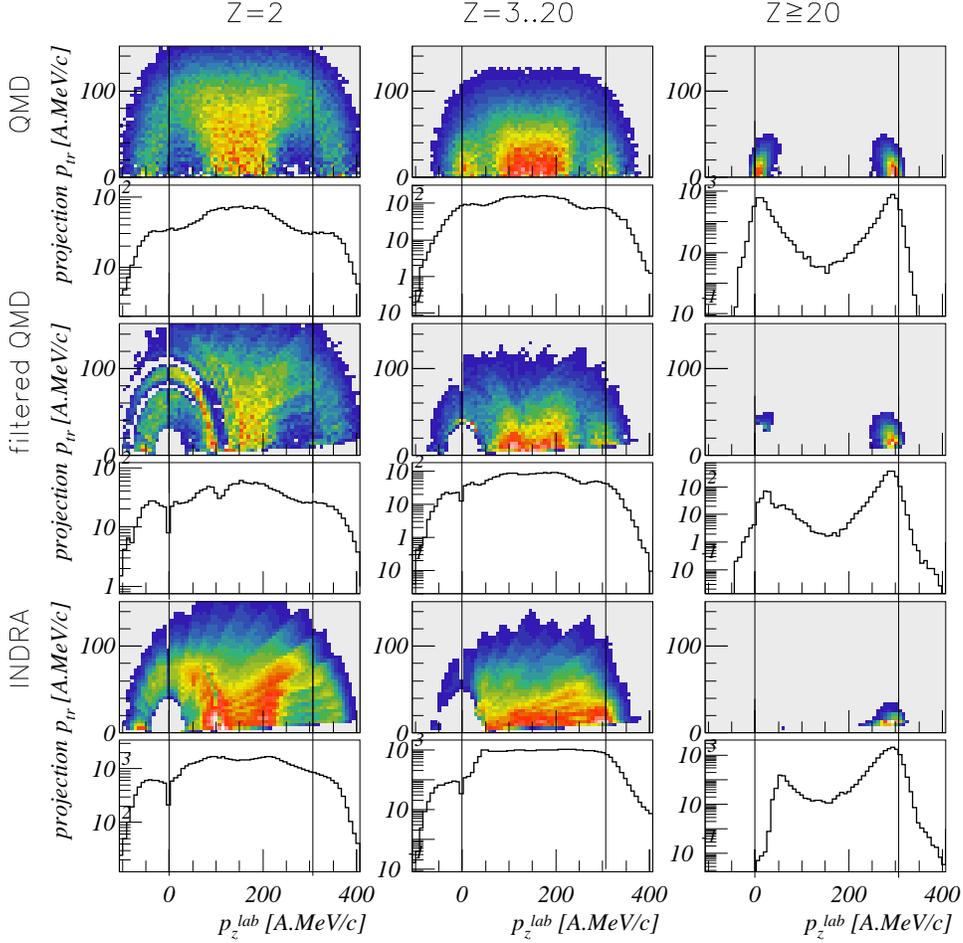}
$$
\vspace{-3cm}
\caption{\textit{Invariant cross section plot of all events. We display QMD
(top), filtered QMD (middle) and INDRA (bottom) data, the lines indicate the
initial projectile and the target momenta. The color coding uses a logarithmic scale }}
\label{wirk1}
\end{figure}

Fig.~\ref{wirk1} displays the invariant cross section ${{d\sigma}\over{dp_z p_t dp_t}}$
in the laboratory system
for three different charge bins (~Z=2, $3 \le Z \le 20$ and $Z > 20$) as
well as the projection onto the $p_z$ axis. We display this quantity for
the unfiltered QMD events, for the filtered QMD events and for the INDRA data.
First of all, we see the importance of the filter. It changes the distribution
in a quite
drastic way.  We see also that the coarse features of the filtered
QMD events are very similar to the INDRA data.  There are, however, several
discrepancies which have to be discussed in view of the event selection which is
used later and in view of the comparison between theory and experiment.

The filter underestimates the blind zone of the detector around the target.
This is true for the Z = 2 particles but even more pronounced for the
intermediate mass fragments (IMF's).
We observe in addition two strong cuts of the
filter for Z =2 particles around $p \approx 100A\  MeV/c$  and $p\approx 80A\ MeV/c$
which are not present in the data on the same level. 

\begin{figure}[h]
\vspace{-3cm}
\epsfxsize=16.cm
$$
\epsfbox{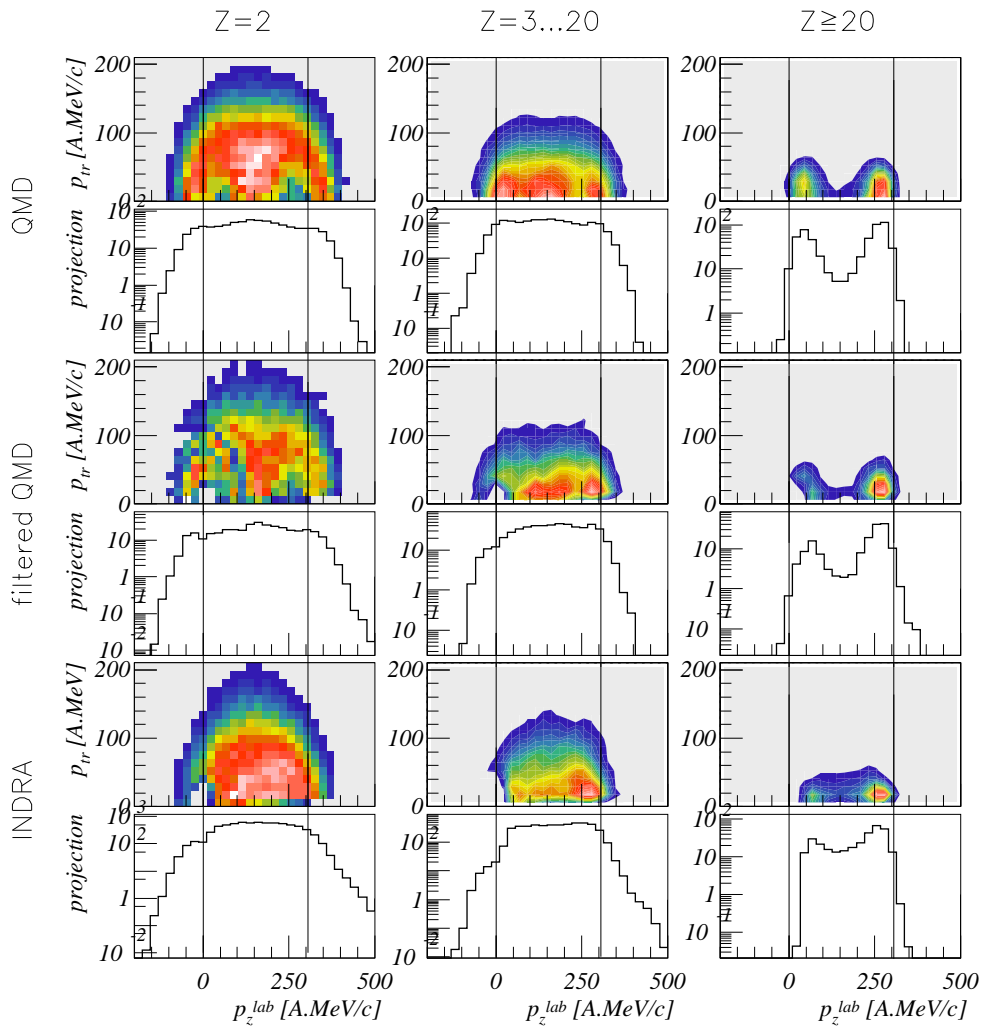}
$$
\vspace{-3cm}
\caption{\textit{Invariant cross section for central events
($E_{trans} \geq 450$ MeV). We display QMD
(top), filtered QMD (middle) and INDRA (bottom) data, the lines indicate the
initial projectile and target momenta. }}
\label{wirkz}
\end{figure}
Whereas this observation points most probably to problems in the filter 
routine, the differences between filtered QMD results and INDRA 
data 
for the heavy fragments $Z \ge 20$  indicate a deficiency of the QMD program.
We observe QMD
events with large projectile and target remnants, in
contradiction to the data. This is due to the fact that these fragments
have gained a transverse in-plane momentum of about 50A MeV/c which is too large as
compared to the data. 
For the INDRA events the transverse momentum gain
of these heavy remnants is slightly smaller and hence
they disappear in the beam pipe. The difference is tiny as far as the 
energy is concerned (the additional transverse energy is less than 1 MeV)
but has important consequences.
We expect for the QMD simulations too 
many events with large projectile and/or target remnants.

The overall structure of the simulated events is rather similar
to that of the INDRA data and hence it is meaningful to proceed further and 
to select central events. As discussed in the next section this is done by
the requirement that the total transverse
energy of the light charged particles is larger than 450 MeV. Before we discuss the 
reason for this choice it is useful to investigate the influence of the filter on
this small subset of events ( cross section $\approx 300 mb$)

Therefore, in fig.~\ref{wirkz} we display the same quantities  as in fig.~\ref{wirk1}
but for central events
only. We see that also for central events the above mentioned discrepancies
between filtered simulations and INDRA data persist. In the target region
too many heavy fragments are observed as compared 
to the data, nevertheless the global event structure is again very similar.


\section{Event selection and impact parameter estimation} \label{impact}

It is the purpose of this article to study central collisions 
of the system Xe(50A MeV)+Sn which has recently been measured by the INDRA 
collaboration. The central collisions are the most interesting ones because
they yield the largest number of fragments and they are those for 
which the system may come to thermal equilibrium.

How can central collisions be identified?
An important property of the INDRA detector is its high efficiency for light
charged (Z=1,2) particles (LCP) independent of the type of the reaction mechanism.
As observed in ref. \cite{LUKA} the
total transverse kinetic energy of the light charged particles (LCP's)  
serves as an indicator for the 
centrality of the reaction. We hence consider the transverse 
energy of all light charged particles 
\beq E_{trans} = \sum_{Z=1,2}E_i \sin^2{\Theta_i} \eeq
as a measure of the impact parameter in QMD events as well as in experiment. 

\begin{figure}[h]
\vspace{-.5cm}
\epsfxsize=15.cm
$$
\epsfbox{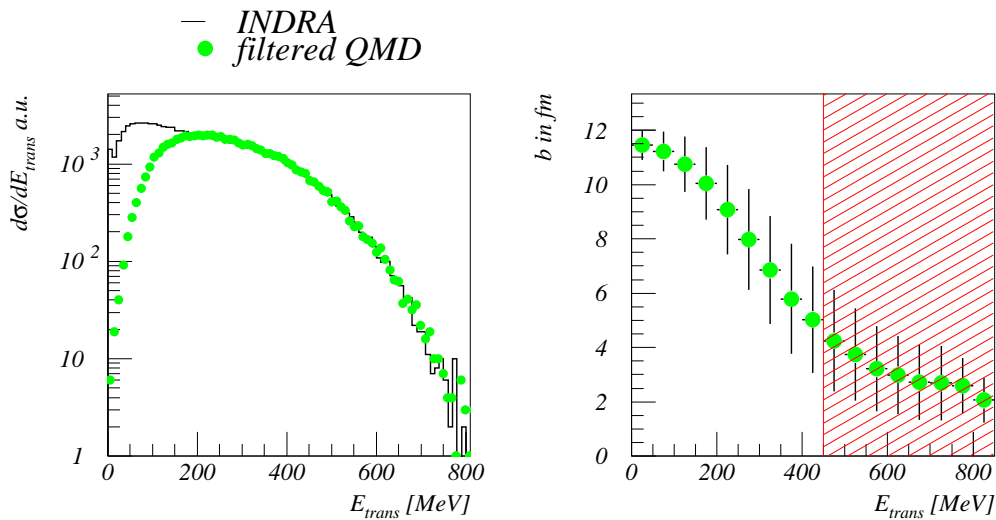}
$$
\vspace{-1.5cm}
\caption{\textit{Transverse energy spectra for INDRA and filtered 
QMD events (left) and the correlation between the transverse energy and the
impact parameter for filtered QMD calculations (right). The error bars represent the
standard deviation. The spectra are normalized on the maximum of the QMD
spectrum.}}
\label{etrb}
\end{figure}

The comparison of theory with experiment has to be done in two steps. 
First we have to make sure that QMD
reproduces the measured transverse energy distribution of light 
charged particles. If this is the case, the same cut in $E_{trans}$ 
selects the same centrality in theory and in experiment. 
In figure \ref{etrb}, left hand side, 
we present the transverse energy spectra for 
QMD simulations and for the INDRA data.
We find that the QMD model reproduces the
cross section $\frac{d^2\sigma}{dE_{trans}}$ quite reasonably, 
(thus the single particle  dynamics is well  described in QMD) and can be
used to characterize the events.
For small transverse energies the deviation is due to the limitation of the calculations
to $b \le 12$. This observation allows already to draw the
first conclusion. It points to the fact that the in medium
nucleon nucleon cross section is not very different from its free
counterpart. Even a mild change on the 20\% level of the cross section 
would have given rise
to a quite different stopping and hence to a quite different transverse 
energy of the LCP's.

One should, however, add a word of warning. QMD deals with effective charges and
there is no unique prescription how to transform clusters consisting of nucleons
with an effective charge to real fragments. Here we have identified a cluster
containing A effective charge nucleons with the most
stable nucleus of a given mass A. This procedure may violate charge 
conservation. A prescription which conserves strictly the total charge
\cite{tirau} yields
a slightly different charge distribution of fragments and consequently a
small change of the total transverse energy of light charged particles.

In figure~\ref{etrb}, right hand side, the correlation between the impact 
parameter and
the transverse energy, as observed in QMD, is displayed. As can be seen, 
the total transverse energy of the
light particles permits a classification of the events according to their 
centrality up to an energy of about 600 MeV.

Being 
interested in central collisions, we will use events with
$E_{trans}\ge 450$~MeV. In 
our simulations the average impact parameter for this choice is $3.6~fm$.

\begin{figure}[h]
\vspace{-2cm}
\epsfxsize=15.cm
$$
\epsfbox{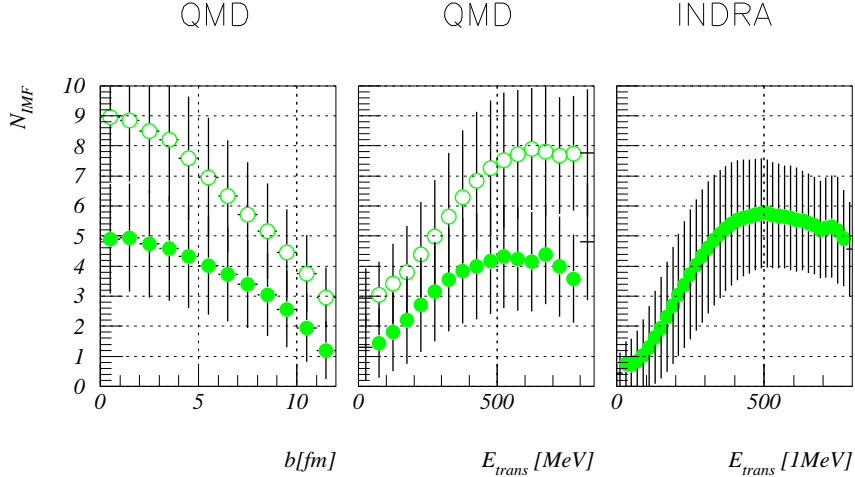}
$$
\vspace{-2.cm}
\caption{\textit{Number of intermediate mass fragments observed as a function
of the impact parameter (lhs) and of the total transverse energy of light
particles (middle). The open circles represent the QMD simulations before being filtered,
the solid circles the filtered QMD data. On the right hand side we plot the
corresponding experimental results. The vertical bars mark the rms deviation.}}
\label{imfeb}
\end{figure}

Can this impact parameter dependence of the transverse energy be verified 
by experimental data? For
this purpose we plot in fig.~\ref{imfeb} the IMF multiplicity as observed in
QMD as a function of the impact parameter. We observe an increase of the
multiplicity with increasing centrality. Despite of the large acceptance of the
INDRA detector we loose about 40\% of the fragments by applying the filter, mostly due 
to the small energy of midrapidity fragments in QMD (section \ref{en2}).

The form of the spectra, which is very similar 
for QMD and INDRA data, is not changed by the filter, 
a reassuring fact for our event selection.

The monotonic dependence of the transverse energy and the fragment multiplicity
on the impact parameter allows to eliminate the
impact parameter in displaying the IMF multiplicity as a function of the
transverse energy. It is displayed in the middle and the right hand side of 
fig. 4.
We see a very good agreement of the form between theory and experiment. The
absolute value of the fragment multiplicity in the simulation is about 
$30\%$ too low, a consequence of the too many accepted events with large 
projectile/target remnants.

There exist other criteria to select the central events. It has been proposed 
\cite{marie,sal} to use the event shape as a selection criteria and the
experimental data have been analyzed using this criterion. We will discuss this 
alternative proposition in chapter \ref{bla}.

\section{single particle spectra}\label{sipa}

\begin{figure}[h]
\epsfxsize=15.cm
\vspace{-1cm}
$$
\epsfbox{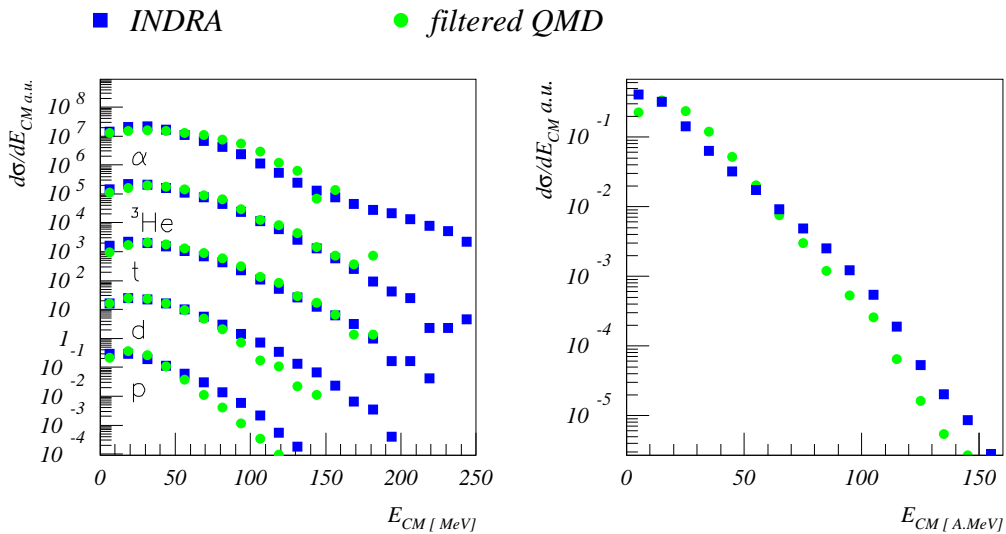}
$$
\vspace{-2.5cm}
\caption{\textit{Angle integrated energy spectra of light charged 
particles (lhs) and 
"protonlike" particles (rhs) for central collisions ($E_{trans} \geq
450$ MeV). The spectra are normalized on the surface}}
\label{spli}
\end{figure}

We saw in fig. \ref{etrb} that the QMD calculations reproduce the
total transverse kinetic energy of the light particles quite well. Now we will
concentrate on central
collisions and discuss whether this good agreement holds also for the spectra 
of light
particles. In figure \ref{spli} we plot the kinetic energy spectra for different
light charged particles. In the QMD model the number of these light particles 
is not very well reproduced. If one analyzes the binding energy of the clusters 
for the Hamiltonian described in section II we find a binding energy which
follows the Weizs\"acker mass formula. The experimental binding energy of light 
clusters
deviates considerably from the value predicted by this formula. Hence loosely
bound clusters like $^3He$ are overpredicted whereas strongly bound clusters
like $^4He$ are underpredicted. Therefore it is useless to compare the absolute
number of these light fragments. However, it makes sense to compare the slope
of their kinetic energy spectra which carries information about the phase space
distribution of the nucleons at the point of their formation. This information
should be rather independent of the observation that in QMD simulations 
too many (too few) of the light fragments are destroyed if their binding energy 
in QMD is too low (too large) as compared to the
real binding energy. The slope of the light charged particles is quite well
reproduced. Deviations are observed for protons and to a lesser extend for deuterons at
high kinetic energy.

One can eliminate the problem with the absolute number of light charged
particles by introducing a spectra for "protonlike" particles by giving each
proton, bound in a fragment of mass A ($A < 4$), the energy $E_{frag}/A$. The 
spectra of
"protonlike" particles are displayed on the right hand side of fig.~\ref{spli}.
The mean kinetic energy /slope of the "protonlike" spectra is 20 MeV /10.5 MeV 
for the filtered QMD simulations and 16 MeV /12.2 MeV for the INDRA data.


\section{Charge-velocity correlations, charge and multiplicity 
distributions of IMF's}\label{zvmulz}
\begin{figure}[h]
\vspace{-2cm}
\epsfxsize=15.cm
$$
\epsfbox{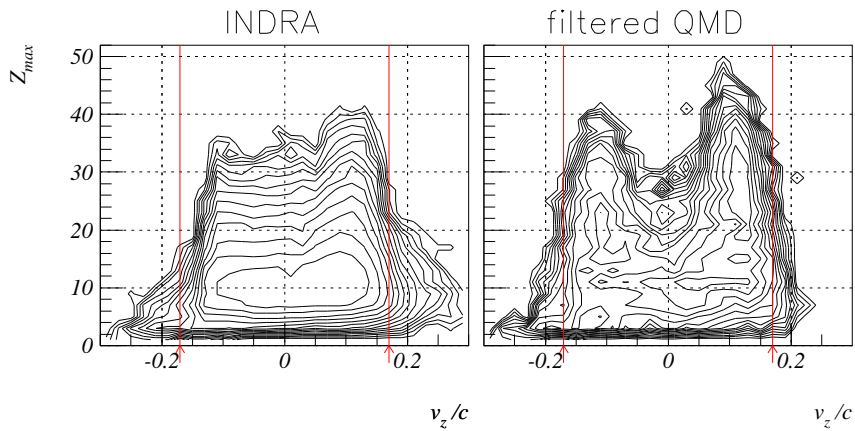}
$$
\vspace{-2cm}
\caption{\textit{Central Collisions ($E_{trans}\geq450$ MeV). Charge of the two
biggest fragments observed as a function of the parallel velocity in the center
of mass frame.}}
\label{zvz}
\end{figure}

As already indicated in figs. \ref{wirk1} and \ref{wirkz} we observe
a binary event structure 
in the INDRA data as well as in QMD simulations even for the most central
collisions. This fact becomes even more pronounced
if one displays (in fig.~\ref{zvz}) ${d\sigma \over dZ_{max} dv_z}$. 
For fragment emission from a statistically emitting source at rest in the center
of mass one would expect
the maximum of the distribution at midrapidity, clearly in contradistinction to
experiment. 

The biggest fragments have lost about 35\% of their initial
velocity (marked by a line) in the data as well as in the simulations. This is remarkable:
in central collisions at lower energies as well as in central collisions
at higher energies we observe almost globally thermalized (sub)systems. At
low energies the effective potential interaction among the nucleons, given by
the real part of the Brueckner G-Matrix, is sufficiently strong to decelerate
projectile and target in order to equilibrate and to form
a compound nucleus. At higher energies the collisions thermalize the
participants
and we find in central collisions a fireball. Here, at intermediate energies, neither the
effective interaction is sufficiently strong to stop projectile and target
nor are the collisions sufficiently frequent because most of them
are Pauli suppressed. As a consequence, we observe in this energy domain a minimum 
of the stopping power of
nuclear matter. The present experiment confirms for the first time this
theoretical prediction. 
The enhancement of QMD events close to projectile and target velocity, amplified by
the logarithmic representation, is another time a consequence of the 
enhanced in-plane transverse momentum in QMD.

\begin{figure}[h]
\vspace{-.5cm}
\epsfxsize=15.cm
$$
\epsfbox{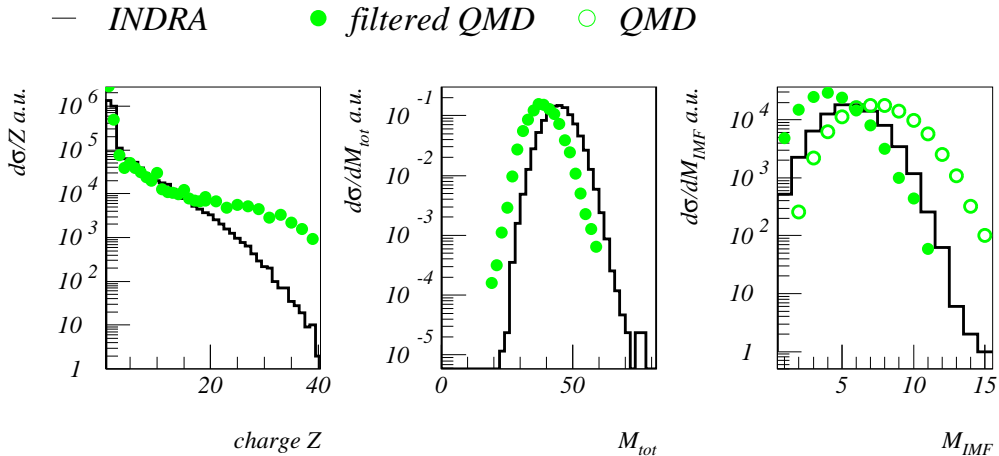}
$$
\vspace{-1.5cm}
\caption{\textit{Central Collisions ($E_{trans}\geq450$ MeV). Charge distribution, 
total
 "protonlike"  and IMF multiplicity 
distributions for INDRA data 
and filtered QMD calculations.}}
\label{multzen}
\end{figure}

The charge distributions, the total
multiplicity of "protonlike" particles  and the IMF multiplicity for the INDRA data and the QMD simulations
(central collisions, $E_{trans}\geq450~MeV$) are shown in figure~\ref{multzen}. 

We find that QMD reproduces well the charge distribution
of the fragments up to $Z\approx15$ but overestimates the production of big
fragments as already discussed. This influences the total multiplicity 
because events with large fragments have a lower multiplicity. 
The IMF multiplicity distribution
(figure~\ref{multzen}, right) gives an average value of 6.6 IMF's observed by 
INDRA, 5.3 IMF's for filtered QMD and 7.5 IMF's for unfiltered QMD. 
The fraction of 
events with a low IMF multiplicity is overestimated in QMD 
which is again a consequence of the too many events
with large fragments which we observe in QMD for this event selection.


\section{Angular distributions} \label{angd}

We come now to the kinematical variables. These, as mentioned in the
introduction, are the key observables for the question 
whether the system or a subsystem
comes close to a statistical equilibrium. Statistical models have been 
very successfully applied to multifragmentation reactions to describe 
multiplicity
distributions, fragment correlations and charge distributions and hence
quantities related to the chemical potential \cite{bon95,bot}. The dynamical
variables, like the kinetic energy spectra or the angular distributions,
available up to now, had not been measured over a sufficient wide energy range in order
to allow for a detailed comparison. With the present data this situation has
changed. Hence we can address the question whether the dynamical observables
show statistical features.

\begin{figure}[h]
\vspace{-2.5cm}
\epsfxsize=15.cm
$$
\epsfbox{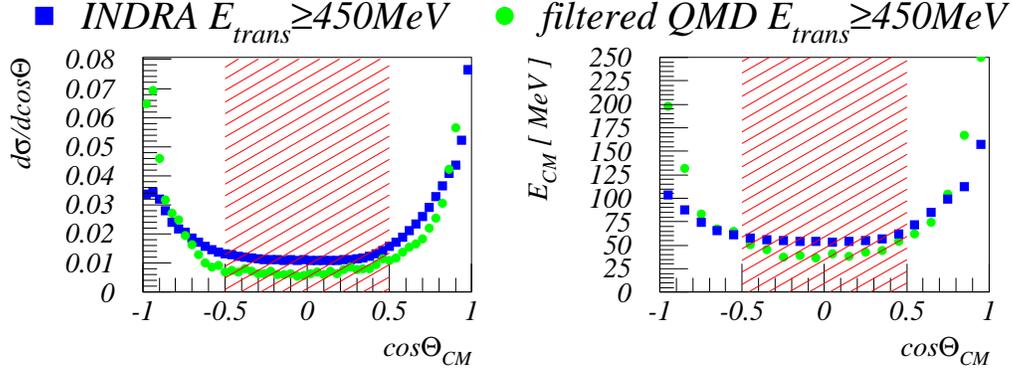}
$$
\vspace{-2cm}
\caption{\textit{Central Collisions ($E_{trans}\geq450$ MeV). Angular 
distribution for IMF's 
and kinetic energy of the IMF's as a function of $\cos\theta$.}
}
\label{ang}
\end{figure}

The angular distribution of IMF's is presented in fig.~\ref{ang}.
We observe a flat distribution for $60^o\le\theta_{CM}\le120^o$ and
a strong increase of the yield in forward and backward direction for both,
filtered QMD events and experiment. Also the average kinetic energy 
is almost constant for $60^o\le\theta_{CM}\le120^o$ and shows as well an 
increase for angles approaching the beam axis. There the 
binary character of the reaction gains influence even for the central collisions. 

This observation has
triggered the conjecture that in these reactions indeed an equilibrated 
subsystem is produced which reveals itself in the midrapidity region 
$60^\circ \le \theta_{CM} \le 120^\circ$. 
In forward ($\theta_{CM} < 60^\circ $) and in backward ($\theta_{CM} > 120^\circ $)
direction it is superimposed by preequilibrium emission.
The quite different behavior makes it meaningful to separate the analysis for
the two zones:
\begin{itemize}
\item the {\bf{midrapidity zone}} which corresponds to fragments emitted in
$60^o\le\theta_{CM}\le120^o$,
\item the {\bf{forward/backward zone}} which corresponds to fragments emitted in

$\theta_{CM}<60^o,\theta_{CM}>120^o$
\end{itemize}

\section{Emission at midrapidity}\label{mid}

\subsection{Charge and IMF multiplicity distributions}\label{mulso}

\begin{figure}[h]
\vspace{-1.cm}
\epsfxsize=15.cm
$$
\epsfbox{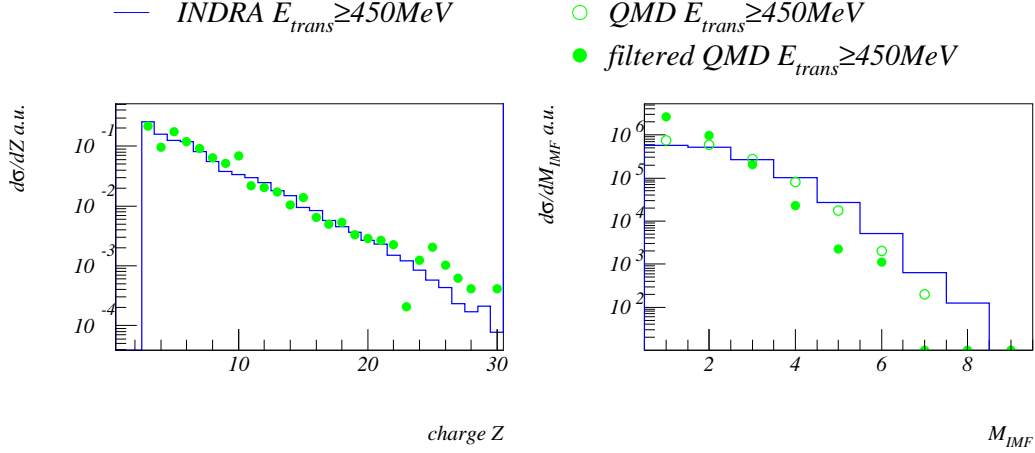}
$$
\vspace{-1.cm}
\caption{\textit{Central Collisions ($E_{trans}\geq450 MeV$). Charge (Z$\geq$3)
and IMF multiplicity distribution for the 
emission at $60^o\le\theta_{CM}\le120^o$ for INDRA data and QMD calculations. }}
\label{multsou}
\end{figure}

The charge yield and the IMF multiplicity distribution of fragments emitted in
the midrapidity zone are plotted in fig.~\ref{multsou}. We find a good 
agreement between the INDRA results and QMD data. The multiplicity distribution
on the right hand side is
once again spoiled by the large fragment events.
In this plot the distributions are normalized on the area.

\begin{figure}[h]
\vspace{-1.cm}
\epsfxsize=10.cm
$$
\epsfbox{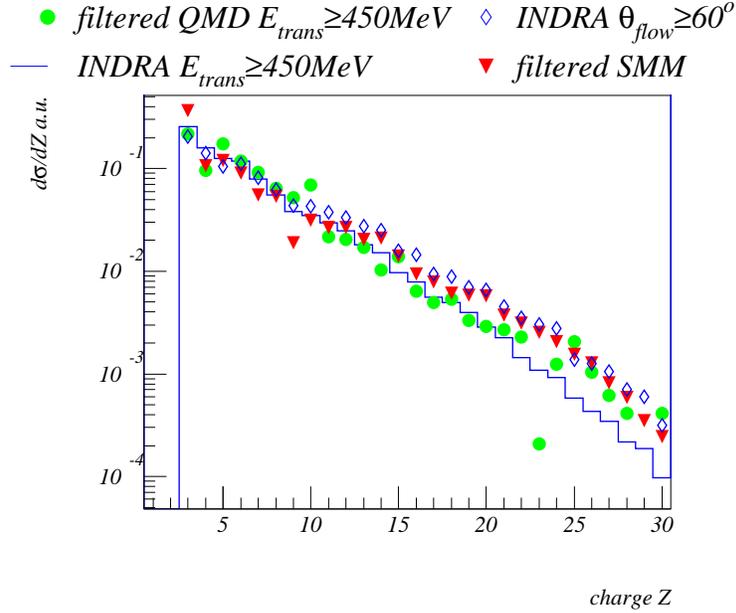}
$$
\vspace{-1.cm}
\caption{\textit{ 
Charge ($Z \ge 3$) distribution of INDRA  for the events with $\theta_{flow} \geq
60^o$ and for events with $E_{trans}\geq450 MeV$ as compared to   
QMD and SMM calculations.}}
\label{zdi}
\end{figure}

If we employ
a rather different event selection criterion, $\theta_{flow}  \geq 60^o$,
discussed in section \ref{bla}, we observe
almost the same charge yield as can be seen in fig. \ref{zdi}.
In ref. \cite{sal} the charge distribution 
for  $\theta_{flow}  \geq 60^o$
has been compared with the prediction of a statistical model calculation
using the SMM program \cite{bon95,bot}. The good agreement 
can be seen as well in fig.~\ref{zdi}. Note that the source sizes
varies, for INDRA ($\theta_{flow}\ge 60^o$) as well as SMM which both correspond to a
single source, the source is about a factor of 1.5 
larger than the QMD and INDRA selection with $E_{trans}\ge 450 MeV$.

Thus at midrapidity the charge distribution is well described by two almost opposite
models for heavy ion collisions, the SMM, which starts out with the assumption
that the system is in a global thermal equilibrium and the QMD which predicts,
as we will see later, that the fragmentation is of dynamical origin. Hence the
charge distribution at midrapidity is not sensitive to the reaction mechanism.


\subsection{Azimuthal distribution in the event plane}

\begin{figure}[h]
\vspace{-3cm}
\epsfxsize=15.cm
$$
\epsfbox{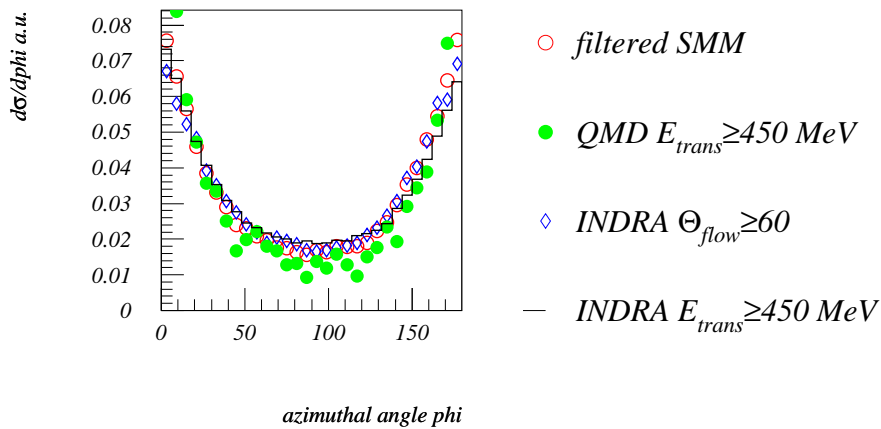}
$$
\vspace{-1.5cm}
\caption{\textit{ Azimuthal distribution of fragments
for the events with $\theta_{flow} \geq
60^o$ and for events with $E_{trans}\geq450 MeV$. We compare the INDRA results
with QMD and SMM calculations }}
\label{azim}
\end{figure}

Another variable for which statistical models make a very definite prediction
is the azimuthal distribution of fragments. In order to determine the
azimuthal distribution of QMD and INDRA events we have first to define an event
plane with respect to which the azimuthal angle is measured. The event plane
is defined by the beam axis and the largest eigenvector of the momentum tensor
(equation (\ref{mote}))
for the filtered QMD and for the INDRA data.
Using this definition of the event plane
we observe the azimuthal distribution displayed in fig.~\ref{azim}.

We see that in QMD and INDRA the
fragments have a strong preference for being emitted in the event plane. 
Even if SMM does not have an event plane, we have to treat it in the same way
as the QMD and INDRA data. Applying the routine to SMM, \cite{bon95,bot} we get
 surprisingly the
same result: a preferred emission at the "reaction plane" and not as expected 
an
isotropic emission.  For low fragment multiplicities the
diagonalisation of the momentum tensor produces always
a preferred  emission direction (an autocorrelation), even in case of a 
statistical emission.
Hence the azimuthal distribution of fragments is another observable 
which does not allow to distinguish between thermal and dynamical emission 
of fragments.


\subsection{Energy spectra and temperatures} \label{en2}

\begin{figure}[h]
\vspace{-2cm}
\epsfxsize=10.cm
$$
\epsfbox{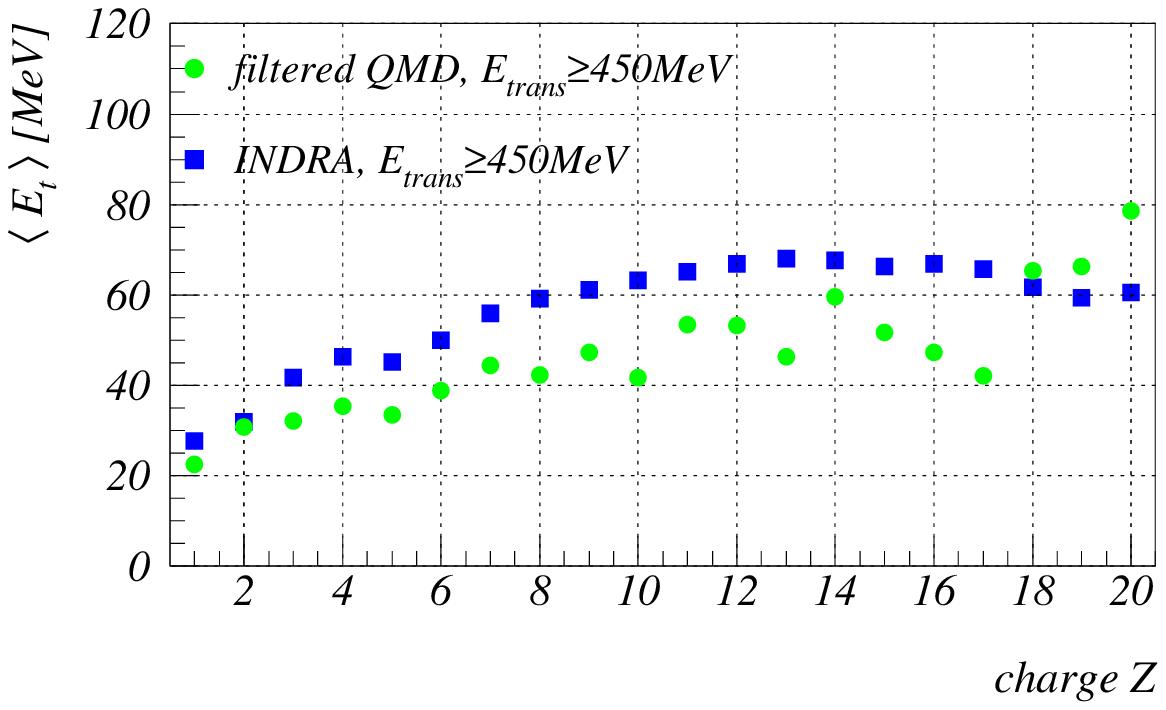}
$$
\vspace{-3cm}
\epsfxsize=10.cm
$$
\epsfbox{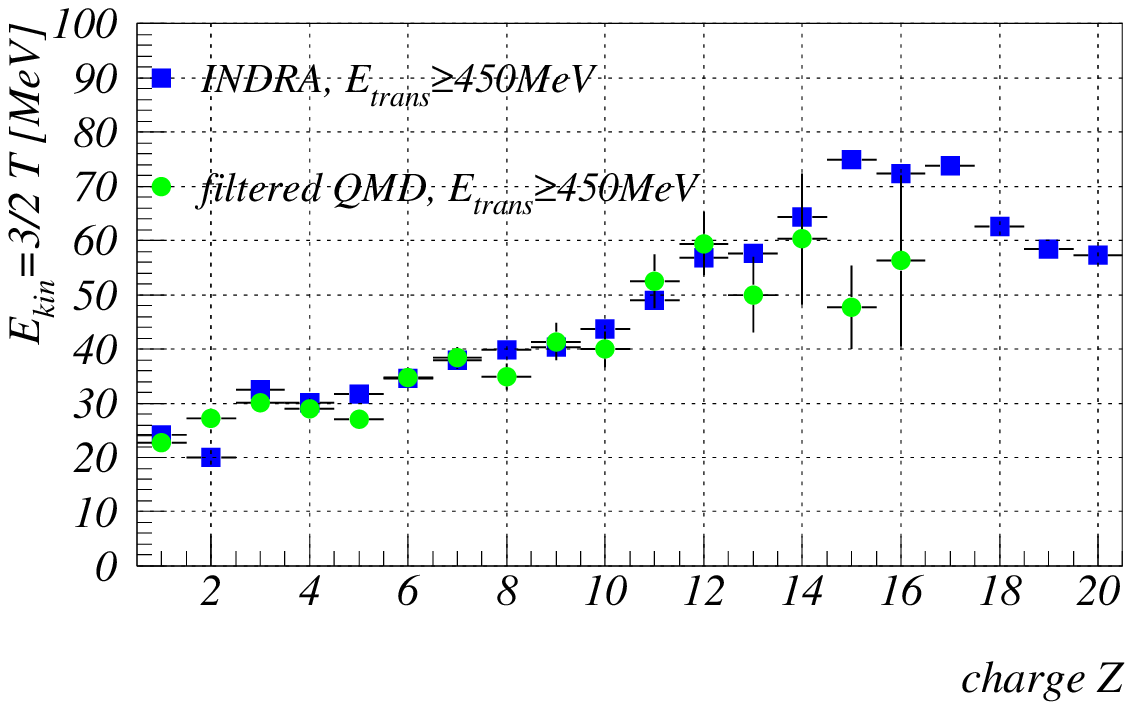}
$$
\vspace{-1cm}
\caption{\textit{Top: Average transverse kinetic kinetic energy with respect to the beam
axis as a function of the charge for fragments 
emitted in $60^o\le\theta_{CM}\le120^o$.  Bottom: Slope of the high energy tail of 
the energy distribution. We display INDRA
data and QMD calculations. Error bars are suppressed.}}
\label{ekmid}
\end{figure}

In figure~\ref{ekmid}, top, we display the average transverse kinetic
energy of fragments with respect to the beam direction emitted at 
midrapidity for the INDRA data and for QMD 
calculations.
We observe a linear rise up to a charge of  twelve, for bigger fragments
where the Coulomb interaction between fragment and system becomes more 
important, the kinetic energy is independent of the fragment size. 
In the case of an emission from a pure thermal source one expects besides the 
modification due to the Coulomb repulsion \cite{sal} the average kinetic 
energy of the fragments to be independent of the fragment mass.
The Coulomb repulsion is not sufficient to explain the increase of the energy with the
fragment mass, therefore the experimental data  contradict to a pure thermal
emission scenario. However, assuming a radial flow proportional to the fragment
mass this can be cured.

This linear increase of the kinetic energy is reproduced by
the QMD data, the absolute value of the energy is, however, underestimated.  
The reason will be discussed later.
In fig.~\ref{ekmid},
bottom, we display a fit to the exponential tail of the fragment kinetic energy 
spectra.
For thermal spectra this slope, being the temperature, has to be constant.
We observe a slight increase of the slope as a function of the fragment charge 
and a good agreement between the INDRA data and
the QMD simulations. 
For big fragments we have too little statistics to allow for a fit of the slope
of the energy spectra obtained in QMD simulations. 
The slope of the INDRA
data increases by almost a factor of 2 between Z=10 and Z = 15. For this
increase there is no remedy in a thermal or a statistical model. Hence we
encounter for the first time in the analysis of the midrapidity fragments
an observable which manifests that the system is not in thermal equilibrium.
Also the absolute value of the slope could  hardly be associated with a
temperature. The binding energy of nuclei is of the order of 8 MeV per nucleon.
If the temperature is substantially higher than this value we do not expect 
that fragments survive. On the contrary, in the fast fragmentation model
\cite{gol}, we expect a slope of about 3/5 $E_F$ which
corresponds to the observed value for light fragments.

\begin{figure}[h]
\vspace{-1cm}
\epsfxsize=15.cm
$$
\epsfbox{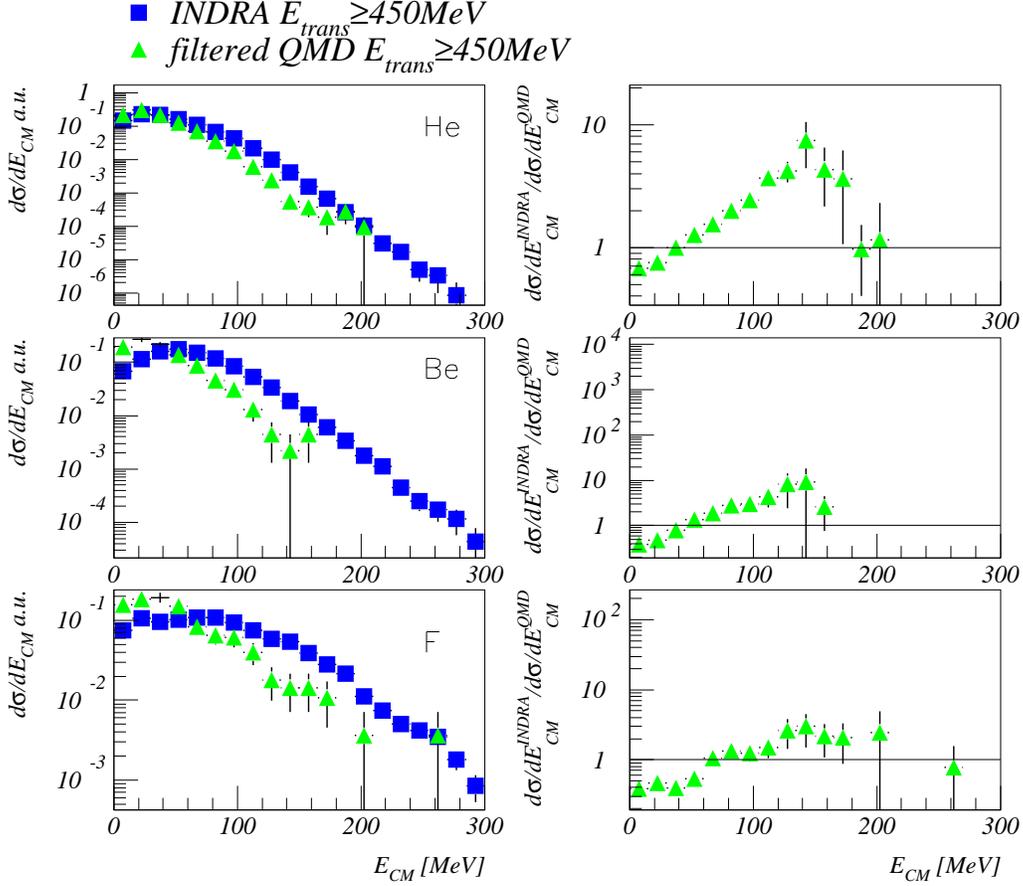}
$$
\vspace{-1cm}
\caption{\textit{Emission at $60^o\le\theta_{CM}\le120^o$ Energy spectra: we compare the
 QMD and INDRA data. On the right hand side the
spectra are displayed, on the left hand side the surprisal analysis.}}
\label{spsouq}
\end{figure}

As the emission of fragments at midrapidity is a rare process which is
logarithmically suppressed for large charges we do not have the statistics 
to compare in detail the kinetic energy spectra of fragments with a charge
larger than 12. For higher charges the fluctuations 
render the analysis meaningless, for the slopes as well as for the spectra.  
The spectra for selected charges are presented in fig. \ref{spsouq}. As already
mentioned, the energy of the fragments in QMD simulations is too small as compared to the INDRA 
data. This points clearly to
a caveat of the simulation program. As explained in section II the effective
range of the nuclear interaction in QMD is too large as compared to
reality. This has the consequence that the repulsive Coulomb force is
compensated by the attractive Yukawa force up to large distances and
hence the gain of kinetic energy due to the Coulomb potential 
is too small. This drawback of the
QMD approach is also responsible for the low multiplicity at midrapidity.
Many QMD fragments are not accepted because they are outside the detector acceptance.
Because the effect of the Coulomb energy becomes more important with increasing charge,
for larger charges the discrepancy between simulation and data becomes more important.

\begin{figure}[h]
\vspace{-2cm}
\epsfxsize=13.cm
$$
\epsfbox{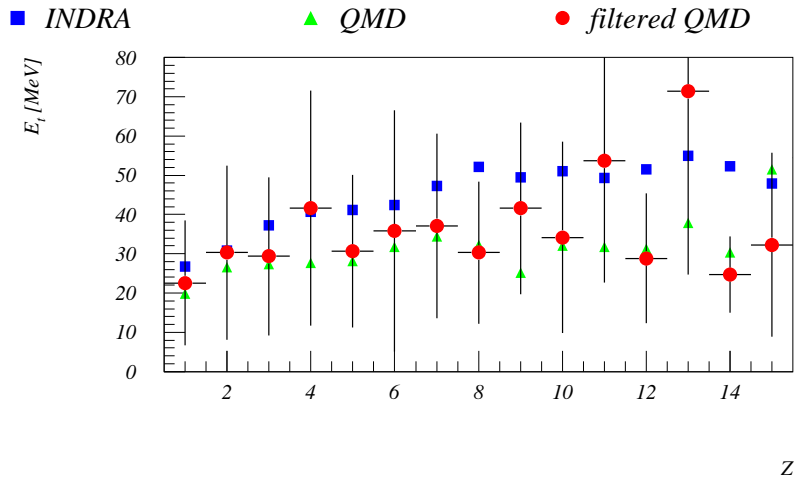}
$$
\vspace{-8cm}
\caption{\textit{Transverse energy (with respect to the largest axis of the momentum 
tensor) of fragments observed in 
$60^o\le\theta_{CM}\le120^o$.
We compare filtered QMD simulations with
experimental data, error bars are represented for filtered QMD only.}}
\label{ettens}
\end{figure}

What is the origin of the increase of the average kinetic
energy with the fragment mass, which is neither expected by a thermal model nor
by an instantaneous fragmentation model? As we will see in these
central collisions the system shows a quite important in-plane flow.
This in-plane flow increases the transverse energy and depends on the
fragment size. In a system which shows transverse flow the beam axis is not
the proper axis of reference for measuring or calculating the transverse
momentum
but has to be replaced by the eigenvector of the momentum tensor 
\beq \label{mote}
P = \left ( \matrix {\sum p_x^ip_x^i & \sum p_x^ip_y^i  &  \sum p_x^ip_z^i \cr
  \sum p_y^ip_x^i & \sum p_y^ip_y^i &   \sum p_y^ip_z^i \cr
   \sum p_z^ip_x^i & \sum p_z^ip_y^i &   \sum p_z^ip_z^i}
   \right)
\eeq
with the largest eigenvalue. 
Therefore we calculate this eigenvector for each INDRA and filtered QMD event 
for the detected IMF's
and determined the transverse energy with respect to this eigenvector. The
analysis is presented in fig. \ref{ettens}. 

We see that in this rotated system the
transverse energy becomes nearly independent of the fragment mass for charges larger
than 3 and the absolute value for the largest fragments is almost half as large 
as compared to the value
obtained with respect to the beam direction.
Thus one has to conclude that the observed increase of the transverse energy with
the fragment mass at midrapidity is nothing but a consequence of the in-plane
flow and has nothing to do with a radial flow. Thus the conjecture to
reconcile the data with statistical emission predictions by introducing a
radial flow does not survive a detailed analysis.  An in-plane flow as a non
equilibrium phenomenon is alien to a statistical model. The absolute value
of the order 40 MeV would give a system "temperature" of 27 MeV, too large to be
considered as the true temperature of the system.

In ref.\cite{ne1} we could show that in QMD calculations the 
kinetic energy observed at midrapidity is a remnant of the initial Fermi-motion
of the nucleons with an additional in-plane flow.
The initial center of mass momentum of all nucleons which finally form 
a fragment is - up to the in-plan flow - almost exactly the same as that 
observed finally. The good agreement between our results and the data
confirm these model predictions of the reaction mechanism.
 
In conclusion, in  our analysis of the midrapidity energy spectra we 
found out
that the mass dependence of the transverse energy is due to the in-plane flow
of the fragments and not due to a radial flow as assumed in a statistical 
interpretation. The
experimental spectra show a charge dependent slope which contradicts to the 
scenario of a thermal freeze out at a given density and temperature. 
The comparison with QMD suffers from the low number of fragments
emitted at midrapidity. Due to an underestimation of the Coulomb barrier
there are many fragments which are below the detector threshold.


\section{Emission of fragments at forward/backward} \label{en1}

We come now to the emission at forward/backward direction. This region is
characterized by a strong dependence of the fragment emission probability 
on the emission angle. It is therefore a region where the system has not even
come close to thermal equilibrium and therefore it presents the challenge
whether simulation programs can predict this highly not equilibrated emission.

\begin{figure}[h]
\vspace{-2.cm}
$$
\epsfxsize=10.cm
\epsfbox{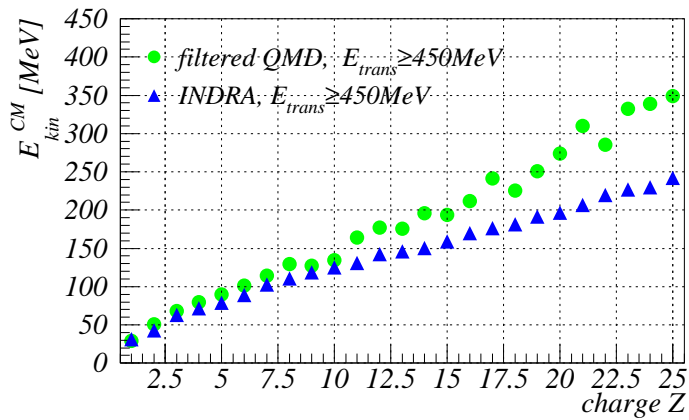}
\vspace{-1cm}
$$
\caption{\textit{Central Collisions ($E_{trans}\geq450$ MeV). Average kinetic energy for
fragments emitted in forward/backward direction, the error bars are suppressed}}
\label{ekinz}
\end{figure}

In forward/backward direction the kinetic energy as a function of the mass shows a
steep linear rise in the QMD simulations as well as in the INDRA data, as
displayed in fig. \ref{ekinz}. For fragments with 
a charge lower than 10 INDRA and QMD agree quantitatively. For larger fragments,
as discussed already, QMD overpredicts the kinetic energy because it generates
large fragments with an unrealistic high in-plane flow. Due to this flow the
fragments are passing the filter routine although in experiment
these big fragments disappear in the beam pipe.

Comparing the
average fragment kinetic energy in forward/backward direction (fig.~\ref{ekinz})
with that at midrapidity (fig.~\ref{ekmid}) we observe a much larger
average energy and the average energy increase in forward/backward direction. This
has to be viewed as a clear manifestation of the projectile and target character 
of these fragments.

\begin{figure}[h]
\vspace{-1cm}
\epsfxsize=15.cm
$$
\epsfbox{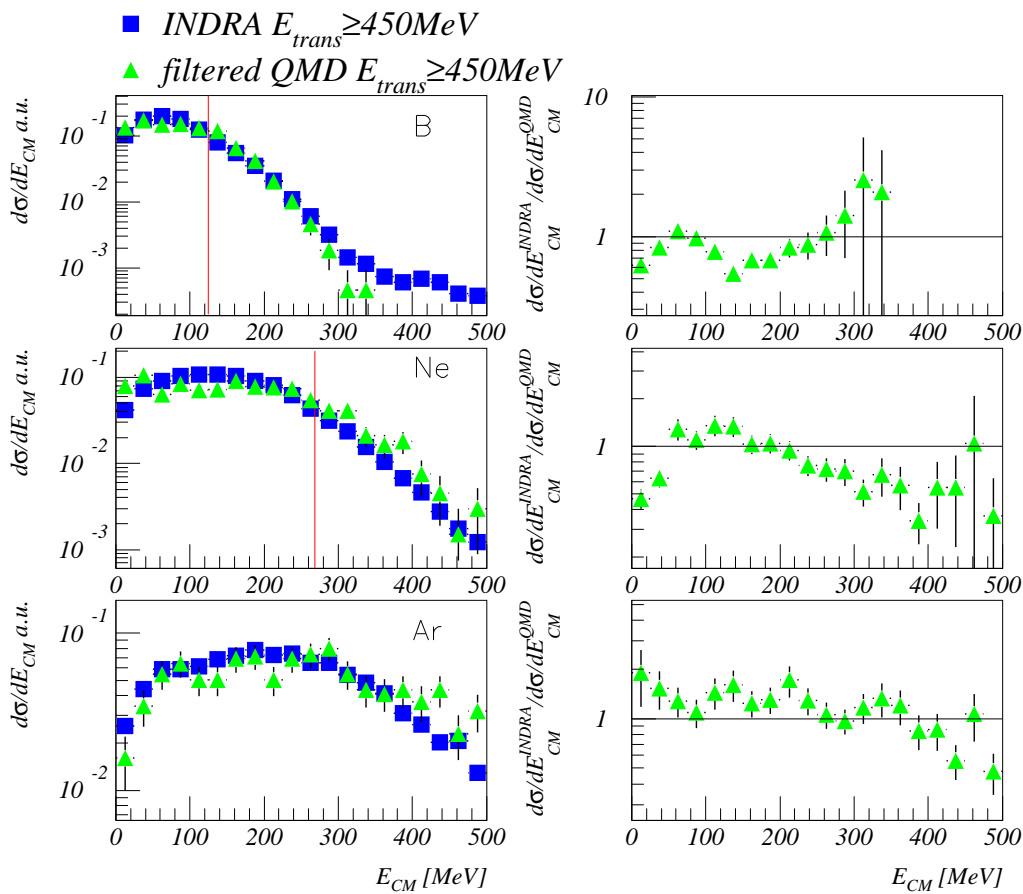}
$$
\vspace{-1cm}
\caption{\textit{Central Collisions ($E_{trans}\geq450$ MeV). Energy spectra for INDRA  
and QMD data in forward/backward
direction in the center of mass frame. On the right hand side we display the logarithm
of the ratio of the two spectra. The vertical lines corresponds to fragments
having the beam velocity.}}
\label{spfb}
\end{figure}

For a more detailed comparison of theory and data we analyze the kinetic energy
spectra for fragments emitted in 
 the forward/backward zone. We display the filtered QMD and INDRA kinetic energy
 spectra   for different charges, Z=5, 10 and 18 in figure
\ref{spfb}.
On the left hand side we show the energy spectra, on the right hand side the 
result of the
surprisal analysis. First we will focus on the INDRA spectra. With increasing mass 
(charge) the maximum
of the spectra is shifted towards higher energies.
The maximum of the distribution in 
forward/backward direction is located already at 
about 80\% of the beam velocity (displayed as a line) and shows therefore clearly 
that most of the
fragments are projectile/target like remnants despite of the fact that in these
central collisions the fragments passed through the whole collision partner.

As can be seen from the spectra the increasing average kinetic energy is not only
due to a shift of the maxima. The slopes change too: the larger the fragment 
the larger the slope. For the case of the Ar fragments one observes already
the overprediction of fragments close to the beam energy in QMD. The low energy
part of the spectra is well reproduced, however.

\section{Other event selection criteria}\label{bla}

As already stated there exist other criteria for finding central
events. Here we follow the
assumption of ref. \cite{sal,lef} that all those events whose flow angle is
larger than $60^o$ can be considered as coming from a central collision. 
We find it useful
to analyze several observables for this criterion.

\begin{figure}[h]
\vspace{-1cm} 
\epsfxsize=16.cm
$$
\epsfbox{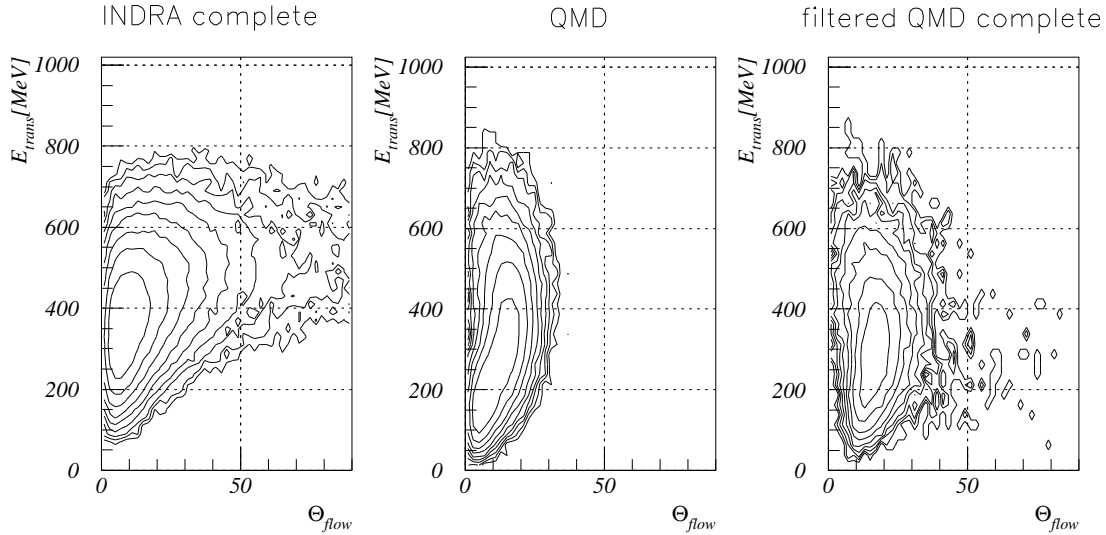}
$$
\vspace{-2cm}
\caption{\textit{Total transverse energy of  light particles versus angle of
flow of the 
fragments, we display complete INDRA (lhs) and QMD (rhs) data. The contour in z is
logarithmic.}}
\label{f1}
\end{figure}

We start our analysis with 
figure~\ref{f1} in which we display the correlation between 
the total transverse energy of light particles as defined in
section \ref{impact} and
the in-plane flow angle of the observed intermediate
mass fragments which serves as criterion in the first conjecture.
On the right
hand side we see the filtered QMD results, in the middle the unfiltered
QMD events and on the left hand side the result of the
experiment \cite{nguy}. The presentation is logarithmic in z.
 In order to be accepted 80\% of the initial charge has to be
detected in experiment as well as in filtered QMD events. We see once more
the consequences of the too large in-plane flow angle of heavy fragments. 
The heavy fragments in the QMD simulation with the large in-plane flow which have already
made problems in analyzing the central events spoil also the spectra obtained under the
completeness criterion. Many of those almost binary events have a small flow and
a small transverse energy $E_{trans}$, i.e. a large impact parameter. These 
events shift the maximum of the cross section to too low transverse energies.

We observe as well that there are too few
QMD events with a large flow angle as compared to experiment. There
about 2\% of complete events show a flow angle larger than $60^o$ whereas in QMD
there are only 0.4\%. The reason for this discrepancy is not easy to determine
as one can see from the figure in the middle where the unfiltered complete 
QMD events are displayed. In the filtered simulation the large flow angle is due to 
the loss of some
fragments, the completeness criterion ensures that these are only small fragments.
In the unfiltered simulations the flow angle of these events is much smaller, 
shows only small
fluctuations and never exceeds $60^o$. 

\begin{figure}[h]
\vspace{-1cm}
\epsfxsize=9.cm
$$
\epsfbox{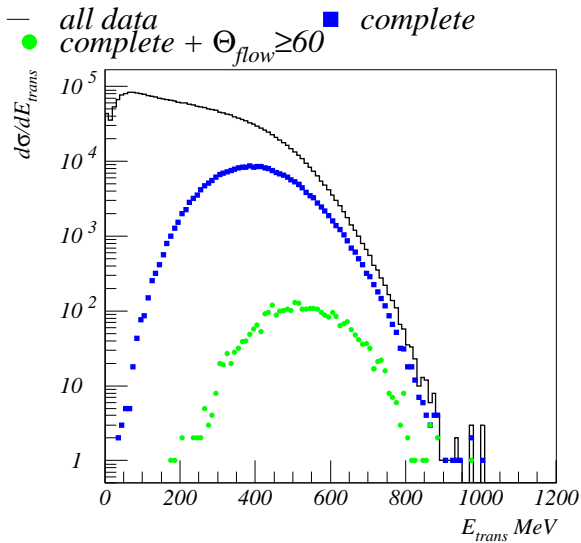}
$$
\vspace{-2cm}
\caption{\textit{INDRA: distribution of $E_{trans}$ for all data, complete events and
events with $\theta_{flow}\ge 60^o$}}
\label{f2}
\end{figure}

There is only a weak correlation between the two centrality
selection criteria as may be seen from fig.~\ref{f2}. There we display the transverse
energy distribution of events with a flow angle larger than $60^o$. We see that
this selection covers a large range of transverse energies extending well below 
$E_{trans} = 450 MeV$ but misses on the other hand a large fraction of events
with $E_{trans} \geq 450 MeV$. As can be inferred from fig. \ref{imfeb} both
criteria cover a quite different range of impact parameters and 
therefore we should not expect that both criteria
give the same results.  

The criterion that the flow angle has to be larger than
$60 ^o$  covers - according to QMD -  a wider impact parameter range.

In view of these facts it is remarkable that most of the observables are very
similar for both criteria. This is even true for the kinetic energy spectra
for fragments emitted at midrapidity.

\section{reaction scenario and conclusions}\label{reac}

The recently measured reaction Xe(50A MeV)+Sn is at the moment the most
complete experiment on multifragmentation.
We have presented a very detailed comparison between the
QMD calculations and the experimental results. We found for most of the
observables a quantitative agreement between the simulations and the experiment.
The QMD simulations has one essential shortcoming which become relevant in
quantitative comparisons on the level which is possible with these new data.
It underpredicts the repulsive Coulomb repulsion and overpredicts the attractive
nuclear interaction due to a range of the nuclear force which is too large 
as compared to reality. As an immediate consequence we observe too small kinetic
energies and larger in-plane flow of the fragments as compared to the
experiment. The too large in-plane flow influences , despite of its small 
value, a whole chain of observables. Heavy fragments which in reality disappear in the 
beam pipe are now observed due to their large in-plane flow. Therefore the mass
yield of QMD disagrees with experiment for charges larger than 30. The events 
with
large remnants have usually a small charged particle multiplicity. Hence the
artificial acceptance of those events lowers the multiplicity of LCP's, however
it does not change the form of IMF observables because IMF's are not produced in
these events. 
 
QMD simulations reveal as the experiment that even central collisions show a
binary character. Thus the stopping power of nuclear matter at this energy is
smaller than at lower energies as well as at higher energies. At this energy we
observe the transition between mean field dominated stopping (as observed at
lower energies yielding compound nucleus formation) and collision dominated
stopping (as observed at higher energies where a fireball is formed). In this
transition region the beam energy is too large for a stopping by the mean field
and too small for a stopping by collisions because most of the collisions
are still Pauli suppressed.

Although in the midrapidity zone the angular distribution is flat and
the energy per particle constant, the matter there is far from equilibrium.
A detailed analysis reveals a strong in-plane flow and a strong azimuthal 
anisotropy. With respect to the largest eigenvector of the momentum tensor
the average transverse kinetic energy of the fragments is constant \cite{ne1}.
The 
linear increase as a function of the fragment mass observed with respect to 
beam axis (fig.~\ref{ekmid}), which has been interpreted as a sign of a 
collective radial flow, 
is merely a consequence of the in-plane flow of fragments.

For several observables we have seen discrepancies to the INDRA data. These discrepancies
are related to shortcomings of the QMD simulation which could be identified. Besides this
we observe a quite good agreement between data and theory and it becomes
useful
to take advantage of the additional coordinate space and time information available
in the QMD simulations. A detailed analysis of the time evolution has been
published elsewhere\cite{ne1}.
It reveals that the final momentum - besides the in-plane flow - of the fragments
is almost exactly that of its progenitor at the beginning of the reaction.
Along the eigenvector of the flow tensor the initial average momentum of 
those nucleons which 
form finally a fragment does change but arrives finally at its initial
values. Hence all changes during the reaction are collective which affects
all fragment nucleons in the same way and is caused by the potential 
interactions. There is no room for a random (thermal) excitation. 
Thus the process is in momentum space very close
to that proposed by Goldhaber 25 years ago: the fragment transverse kinetic energy is a
consequence of the initial Fermi energy. It is a convolution of the momenta of
the entrained nucleons, each of them having the momentum distribution of
nucleon in a Fermi gas. 

Goldhaber assumed that the fragmentation is a fast process. This we cannot
confirm. The fragments decouple from the system only after 200 fm/c.
Therefore the question remains how the fragments pass the surrounding
matter without being destroyed. The two body potential interaction is rather
smooth and acts on all fragment nucleons very similar. Hard two body collisions
may transfer momentum to one of the fragment nucleons what may lead to a
separation of that nucleons from the other fragment nucleons. 
The key quantity to understand this passing trough
is the mean free path. At the energies considered here it is quite large due to
the Pauli suppression of the collision. Therefore in a very simplified model,
which contains however the essential physics, multifragmentation can be viewed
as two lattices, representing the nuclei, passing through each other. Each
lattice site is occupied by a nucleon.
Collisions
transfer momentum, the scattered nucleons escape and leave holes at the 
corresponding lattice sites. If sufficient but not too many holes are created 
the lattices separate into disconnected parts, which are the fragments. If more
collisions occur there are more holes and consequently less fragments, if less
collisions occur we have only connected parts,i.e. remnants of projectile and 
target.
Thus nucleons being finally part of fragments have only suffered from small
momentum transfers during the collision, or, inversely, nucleons which had had a
hard collisions are not part of a fragments. Of course in reality the situation
is much more complicated because the nucleons have momenta but these additional
features do not change the qualitative description. 

This scenario is also the
basis of the success of the percolation model which describes multiplicity
distributions and the charge yield of multifragmentation quite satisfying. 
The detailed investigation of how the fragments pass the surrounding matter 
has been performed in ref. \cite{goss97} long before data became available.
That this reaction scenario is now confirmed by experiment demonstrates that
the simulations of heavy ion reactions at this energy has reached a level which
allows to draw firm conclusions about the underlying reacting mechanism and
allows to understand the evolution of the finite number non equilibrium quantum
systems.   

It is remarkable that statistical models yield - in the limited kinematical
regimes where they can be applied - almost the same value for the different
observables. This has been interpreted in the past as a strong indication
that the system comes to a global equilibrium. This we cannot confirm. In view
of our results this agreement is inconclusive as far as the reaction mechanism
is concerned. 

\end{document}